\documentclass[%
 reprint,
 amsmath,amssymb,
 aps
]{revtex4-1}
\usepackage{mathptmx}
\usepackage{subfigure}
\usepackage{graphicx}% Include figure files
\usepackage{dcolumn}% Align table columns on decimal point
\usepackage{bm}% bold math
\usepackage{hyperref}% add hypertext capabilities

\begin{document}

\preprint{APS/123-QED}

\title{Random walks in time-varying networks with memory}% Force line breaks with \\
%\thanks{A footnote to the article title}%

\author{Bing Wang}
\email{bingbignwang@shu.edu.cn}
\author{Hongjuan Zeng}
\author{Yuexing Han}
\email{han$_$yx@i.shu.edu.cn}
\affiliation{School of Computer Engineering and Science, Shanghai University, Shanghai, P.R. China
}
%\collaboration{MUSO Collaboration}%\noaffiliation
%\collaboration{CLEO Collaboration}%\noaffiliation

\date{\today}% It is always \today, today,
             %  but any date may be explicitly specified

\begin{abstract}
Random walks process on networks plays a fundamental role in understanding the importance of nodes and the similarity of them, which has been widely applied in PageRank, information retrieval, and community detection, etc. Individual's memory has been proved to be important to affect network evolution and dynamical processes unfolding on the network. In this manuscript, we study the random-walk process on extended activity-driven network model by taking account of individual's memory. We analyze how individual's memory affects random-walk process unfolding on the network when the timescales of the processes of the random walk and the network evolution are comparable. Under the constraints of long-time evolution, we derive analytical solutions for the distribution of stationary state $W_a$ and the mean first-passage time~(MFPT) of the random-walk process. We find that, compared with the memoryless activity-driven model, individual's memory enhances the fluctuation of degree distribution, which reduces the capability of gathering walkers for nodes, especially with large activity and delays the mean first-passage time. The results on real networks also support the theoretical analysis with artificial networks.
\end{abstract}

\keywords{Suggested keywords}%Use showkeys class option if keyword
%display desired
\pacs{\textbf{check the PACS codes}}
\maketitle

%\tableofcontents

\section{\label{sec:level1}Introduction}
Random walks on networks describes a diffusion process, which has broadly been applied in ranking systems \cite{gleich2015pagerank}, community detection, \cite{rosvall2008maps} and decision-making \cite{gold2007neural}. According to different rules, random walks in static networks can be divided into classical random walks \cite{lovasz1993random}, self-avoiding walks \cite{domb2009self}, biased random walks \cite{fronczak2009biased}, and quantum walks \cite{venegas2012quantum}. Among them, the classical random walks is widely studied, which describes the process that the walker has no memory for the path and moves to its neighboring nodes with equal probability.

In the early stage of network research, due to the limitations of data collection and storage equipment, a large amount of research work focused on static time-aggregated networks, in which edges between nodes do not change over time~\cite{eames2002modeling,newman2006structure}. However, most of complex systems in nature, society, and technology show temporal characteristics, where the pattern of connections between individuals evolves in time ~\cite{curci2012temporal}. The increasingly accurate marking of temporal data facilitates the description of network structures~\cite{kim2012temporal,kim2015scaling}.

Thus, more attention to random walks in time varying networks has been paid with the help of the activity-driven network model~\cite{perra2012activity}. In the activity-driven model, each node in the network is activated according to the pre-assigned activity which describes the propensity of the node to form connections. Although the model is simple, it describes the characteristics of temporality and degree distribution of real systems.
Unlike annealed and quenched networks, random walk diffusion process is affected by the temporal connectivity patterns between nodes, ~\cite{perra2012random,hoffmann2013random,Alessandretti2017Random,yang2018random,moinet2019random}, which means that walkers can get trapped at temporarily isolated nodes. It shows that nodes with large activity have strong ability to collect walkers and reduce the MFPT \cite{perra2012random}. By taking account of individual's attractiveness, it shows that heterogenous attractiveness limits nodes' ability to collect walkers, especially when attraction and activity are positively correlated~\cite{Alessandretti2017Random}. When links are established by the combination of node's fitness and activity, nontrivial effect has been found on the properties of random-walk process~\cite{yang2018random}. The activity-driven model is further extended by considering burstiness ~\cite{moinet2019random,moinet2015burstiness,ubaldi2017burstiness}, modularity~\cite{nadini2018epidemic}, coupled structures~\cite{lei2016contagion}, and multitype intetactions~\cite{NingNing2019Impacts}. Interaction of nodes in groups of arbitrary numerosity is recently studied~\cite{benson2016higher,grilli2017higher}, which is modelled by simplex complexes~\cite{iacopini2019simplicial,devriendt2019simplex} or hypergraphs ~\cite{bellaachia2013random}.

In real networks, however, edges between nodes are not randomly connected as described in the activity-driven model, but are affected by the non-Markovian effect due to individual's memory ~\cite{sun2014epidemic,kim2015scaling,zino2018modeling}. Individuals tend to interact with people they already know, establishing strong or weak links with them, which can restrain the rumor spreading~\cite{Karsai2014Time}. A reinforcement process encoded with a measurable parameter of memory has been studied in recent work~\cite{ubaldi2016asymptotic}. Limited by the long evolution of the network, memory reduces the threshold of the Susceptible-Infected-Susceptible~(SIS) model and promotes epidemic spreading, which is same for Susceptible-Infected-Recovered~(SIR) dynamics~\cite{tizzani2018epidemic}. In addition, the model of second-order and even higher-order network memory is proposed by defining edge path, which may speed up or reduce the diffusion process and affect the community detection ~\cite{lambiotte2015effect,scholtes2014causality}.

In this manuscript, we investigate random walk process on an extended temporary network based on activity-driven model with individual's memory~\cite{ubaldi2016asymptotic}. This feature of memory accounts for the fact that social interactions are not randomly established but concentrated towards already contacted nodes. We study random walk process unfolding in activity-driven time-varying networks with a parameter $\beta$ tuning the memory strength ~\cite{tizzani2018epidemic}. In the long time limit, we find analytical solutions for $W_a$ and MFPT, respectively. When random-walk process starts after a period of network evolution, the numerical simulation results agree well with theoretical analysis. Compared with the memoryless case, individual's memory increases the fluctuation of degree distribution, thereby reducing the ability of gathering walkers of the nodes with large activity and delaying the MFPT of each node. We then study how memory affects random-walk process in real systems. By comparing random-walk process on the null model with real dataset, we find that individual's memory reduces node's capability of gathering walkers, which is consistent with what is observed in synthetic networks.

The manuscript is organized as follows. In Section~\ref{MODEL}, we introduce the time-varying network model with memory and describe random-walk process. In Section~\ref{AnalyticalResults}, we study the stationary state and the MFPT of the random-walk diffusing on the network model. In Section~\ref{RealWR}, we analyze the strength of individuals' memory in real systems and study the stationary state of the random-walk diffusion on it. Finally, in Section~\ref{Discussion}, we summarize our work.

\section{~\label{MODEL}MODEL}
In activity-driven framework, we define that node's propensity to establish contacts per unit time follows a given power law distribution $F(a)\propto a^{-\gamma}$ with $\varepsilon\leq a_{i}\leq 1$, where $\varepsilon$ is a cutoff value that is chosen to avoid possible divergence of $F(a)$ close to the origin. The dynamics occurs over discrete steps of length $\Delta t$. At each step, with probability $a_{i}\Delta t$, node $i$ becomes active
and connects to $m$ other nodes. When edges between nodes are randomly selected, the degree distribution of the network satisfies the condition $\rho(k)dk \propto F(k)dk$ ~\cite{perra2012activity}.

We demonstrate individual's memory used in this work. Nodes frequently connect with acquaintances, while they rarely contact with new nodes due to the effect of individual's memory. For each active node $i$, which
has already connected $k_{i}(t)$ distinct nodes at time $t$, connects with a new
node with probability $
P_{new,i}(t)=\left[1+k_{i}(t)/c\right]^{-\beta},~\label{eq:rpb}$
while it establishes a connection with a previously contacted node
with complementary probability $P_{old,i}(t)=1-P_{new,i}(t)$~\cite{tizzani2018epidemic}, where the constant $c$ sets an intrinsic value for the number of connections that node $i$ is able to engage in before memory effects become relevant~\cite{ubaldi2016asymptotic}.
The parameter $\beta> 0$ tunes the memory. The larger $\beta$ is, the stronger tie (the larger link-weight) between already connected nodes will be. $\beta =0$ corresponds to the case of no memory. With no loss of generality, we set $c=1$.
\begin{figure}[htbp]
\centering
\includegraphics[width=0.5\textwidth]{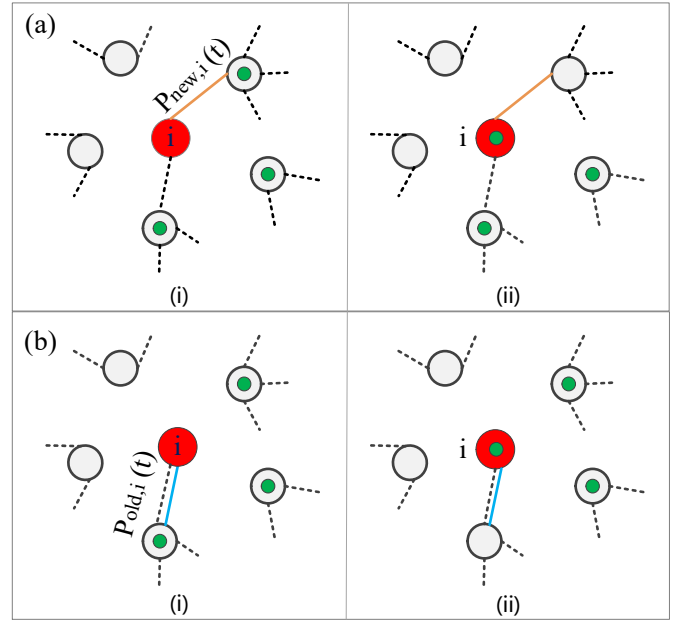}
\caption{\textbf{Random walk in time-varying networks with memory.} The left two panels represent how an active node establishes a link with another node. Each active node, with probability $P_{new(old),i}$ connects with a new~(old) neighbor shown as panel $(a)$ and panel $(b)$, respectively. Walkers are presented as fully green nodes. Active and non-active nodes are shown as red and gray nodes, respectively. The edges between nodes already connected before are shown as grey dotted lines, and current contacts are shown as solid line.}
~\label{figure:1}
\end{figure}

We demonstrate the activity-driven network model with memory as follows:
\begin{itemize}
	\item At each discrete time step $t$, the network $G_{t}$ starts with $N$ disconnected nodes;
	\item With probability $a_{i}\Delta t$, node $i$ generates $m$ links;
    \item With probability $1-P_{new,i}(t)$, nodes $i$ connects with one of the $k_{i}(t)$ previously connected nodes, or with probability $P_{new,i}(t)$, it connects to a new node $j$. Non-active nodes can still receive connections from other active nodes (as shown in Fig.~\ref{figure:1}).
	\item At time $t+\Delta t$, the memory of each node is updated and the process starts over again to generate the network $G_{t+\Delta t}$.
\end{itemize}
$k_{i}(t)$ is the number of different neighbors of node $i$ contacted until time $t$. All the interactions have a constant duration $\triangle t$. Without loss of generality, in the following, we set $\triangle t =1$.

When the network evolves for a long enough time, the number of acquaintance nodes for each node is large enough (i.e., $1\ll k\ll N$), then the probability that the node connects to new nodes is almost negligible. In this case, the networks' connectivity patterns is equivalent to the static network, and the degree of nodes with activity $a$ is $ \bar{k}(a,t)=C(a)t^{1/(1+\beta)}$ ~\cite{ubaldi2016asymptotic}. The perfector $C(a)$ is determined by the condition
$ \frac{C(a)}{1+\beta}=\frac{a}{C^{\beta}(a)}+\int da\frac{F(a)a}{C^{\beta}(a)}.\label{Ca} $ Hereafter, we denote $\langle g\rangle=\int daF(a){g(a)}$
as the average of a function of the activity $g(a)=\frac{a}{C^{\beta}(a)}$ over the network. The degree distribution of the time-varying network with memory is given by $\rho(k)dk \propto F(k^{1+\beta})k^{\beta}dk$ \cite{ubaldi2016asymptotic}.

\section{~\label{AnalyticalResults}Analytical Results}
The probability that the walker stays at node $i$ at time $t$, $P_{i}(t)$, obeys the master equation~\cite{perra2012random}, given by
  \begin{equation}
  \label{eq:1}
  P_{i}(t+\Delta t)=P_{i}(t)\left[1-\sum _{j\ne i}{\Pi }_{i\to j}^{\Delta t}\right]+\sum_{j\ne i}P_{j}(t){\Pi_{j\to i}^{\Delta t}},
  \end{equation}
where ${\Pi_{i\to j}^{\Delta t}}$ is the probability that the walker moves from node $i$ to node $j$ in a time interval $\Delta t$. The first term on the right-hand represents the probability that the walker at node $i$, at time $t$ does not jump to other nodes at time $t+\Delta t$. The second term represents the probability that the walker at neighboring nodes of node $i$, at time $t$, moves to node $i$ at time $t+\Delta t$.

Let us define ${\Omega }_{i\to j}^{\Delta t}$ as the probability that node $i$ becomes active and connects to node $j$, given as follows:
\begin{equation}
\label{eq:2}
{\Omega }_{i\to j}^{\Delta t}=a_{i}m\Delta t\left[\frac{(1-P_{new,i}(t))A_{ij}(t)}{k_{i}(t)}+\frac{P_{new,i}(t)}{N-k_{i}(t)-1}\right].
\end{equation}
The fist term represents that node $i$ activates and selects acquaintance nodes to establish connections. The second term is due to that node $i$ activates and creates a new connection. $A_{ij}(t)$ is the actual adjacency matrix of the network until time $t$, i.e., it is equal to $1$ if node $i$ and node $j$ have been in contact at least once in the past and $0$ otherwise. We can see that both $A_{ij}(t)$ and $k_i(t)$ depend on the evolution time $t$, so the number of walkers distributed as nodes with activity $a$, $W_a$, and MFPT are affected by the starting time of the diffusion.
In this case, the instantaneous degree of node $i$ is $k_{i}=m+\sum_{j}{\Omega_{j\to i}^{\Delta t}}$. Indeed, node $i$ will generate $m$ links and may potentially receive links from other active nodes. The probability that node $j$ is active and connects with node $i$ is instead given by
\begin{equation}
~\label{eq:3}
{\Omega }_{j\to i}^{\Delta t}=a_{j}m\Delta t\left[\frac{(1-P_{new,j}(t))A_{ij}(t)}{k_{j}(t)}+\frac{P_{new,j}(t)}{N-k_{j}(t)-1}\right].
\end{equation}
In this case, the instantaneous degree of node $i$ is $k_{i}=1+\sum_{l\ne j}\Omega_{l\to i}^{\Delta t}$. Here, we assume that when node $j$ is activated, a connection is established to node $i$. The former represents that node $j$ becomes active and connects to node $i$, while the latter is that the activated nodes except node $j$ establish links with node $i$.

Considering the events described by Eq.(\ref{eq:2}) and Eq.(\ref{eq:3}) cannot happen at the same time. According to the rules of random walk, a walker staying at node $i$ randomly jumps to one of its $k_i$ neighboring nodes. Putting them all together, the probability that a random walker moves from node $i$ to one of its neighbors, $\Pi_{i\to j}^{\Delta t}$, can be written as,
\begin{align}
\label{eq:4}
\nonumber{\Pi_{i\to j}^{\Delta t}}&={\Omega_{i\to j}^{\Delta t}\frac{1}{m+\sum_{j}{\Omega_{j\to i}^{\Delta t}}}}+{\Omega_{j\to i}^{\Delta t}\frac{1}{1+\sum_{l\ne j}{\Omega_{l\to i}^{\Delta t}}}}\\\nonumber
&\simeq a_{i}\Delta t\left[\frac{(1-P_{new,i}(t))A_{ij}(t)}{k_{i}(t)}+\frac{P_{new,i}(t)}{N-k_{i}(t)-1}\right]
\\& +a_{j}m\Delta t\left[\frac{(1-P_{new,j}(t))A_{ij}(t)}{k_{j}(t)}+\frac{P_{new,j}(t)}{N-k_{j}(t)-1}\right].
\end{align}
Indeed, a link between node $i$ and node $j$ can be established as a consequence of the activation of node $i$ or $j$. Then, we can write the equation describing the evolution of $P_{i}(t)$ by substituting the expression of the propagator in Eq.(\ref{eq:1}):
  \begin{widetext}
  \begin{align}
  \label{eq:5}
 \nonumber&\frac{\partial P_{i}(t)}{\partial t}=\\
  \nonumber &
 -P_{i}(t)\sum_{j\ne i}a_{i}\left\{\left[\frac{(1-P_{new,i}(t))A_{ij}(t)}{k_{i}(t)}+\frac{P_{new,i}(t)}{N-k_{i}(t)-1}\right]
 +a_{j}m\left[\frac{(1-P_{new,j}(t))A_{ij}(t)}{k_{j}(t)}+\frac{P_{new,j}(t)}{N-k_{j}(t)-1}\right]\right\}\\
 & +\sum_{j\ne i}P_{j}(t)\left\{a_{j}\left[\frac{(1-P_{new,j}(t))A_{ij}(t)}{k_{i}(t)}+\frac{P_{new,j}(t)}{N-k_{j}(t)-1}
 \right]+a_{i}m\left[\frac{(1-P_{new,i}(t))A_{ij}(t)}{k_{j}(t)}+\frac{P_{new,i}(t)}{N-k_{i}(t)-1}\right]\right\}.
  \end{align}
\end{widetext}

When considering the evolution of time-varying networks for a long time, the degree of node $i$ follows $1\ll k_i(t)\ll N$, thus the probability that node $i$ connects with new nodes can be ignored, while the network is still a sparse graph. In this limit case, we replace $N-k_i(t)-1$ with $N$. Considering only the leading terms, Eq.(\ref{eq:5}) can be rewritten as
  \begin{align}
  \label{eq:6}
 \frac{\partial P_{i}(t)}{\partial t}\nonumber&=-P_{i}(t)\sum_{j\neq i}{A_{ij}(t)}\left[\frac{a_{i}}{k_{i}(t)}+\frac{ma_{j}}{k_{j}(t)}\right]
 \\&+\sum_{j\neq i}P_{j}(t){A_{ij}(t)}\left[\frac{a_{j}}{k_{j}(t)}+\frac{ma_{i}}{k_{i}(t)}\right],
  \end{align}
where $g(a_{i})=\frac{a_{i}}{C^{\beta}(a_{i})}$, and $\frac{C(a_{i})}{1+\beta}=\frac{a_{i}}{C^{\beta}(a_{i})}+\int da\frac{F(a)a}{C^{\beta}(a)}.\label{Ca}$

To proceed further study, we perform equivalent analysis of the heterogeneous
mean-field approximation for static networks, that is, we replace the
time-integrated adjacency matrix $A_{ij}(t)$ with its annealed form, i.e.,
$Q_{ij}(t)=(1+\beta)t^{1/(1+\beta)}\left[g(a_{i})+g(a_{j})\right]/N$, which describes the probability that node $i$ and node $j$ have
been in contact in the past \cite{tizzani2018epidemic}. We further replace $k_i(t)$ with $\bar{k}(a_{i},t)=(1+\beta)(g(a_i)+\langle g\rangle)t^{1/(1+\beta)}$, Eq.(\ref{eq:6}) can be written as
\begin{widetext}
  \begin{align}
  \label{eq:7}
 \frac{\partial P_{i}(t)}{\partial t}=-P_{i}(t)\sum_{j\neq i}\left[\frac{a_{i}(g(a_i)+g(a_j))}{g(a_i)+\langle g\rangle}+\frac{ma_{j}(g(a_i)+g(a_j))}{g(a_j)+\langle g\rangle}\right]+\sum_{j\neq i}P_{j}(t)\left[\frac{a_{j}(g(a_i)+g(a_j))}{g(a_j)+\langle g\rangle}+\frac{ma_{i}(g(a_i)+g(a_j))}{g(a_i)+\langle g\rangle}\right],
  \end{align}
\end{widetext}

 We obtain a system level description of the process by grouping nodes in the same activity class $a$, assuming that they are statistically equivalent~\cite{perra2012random}.
Then, we define the number of walkers at a given node of class $a$ at time $t$ as $W_{a}(t)=\left[NF(a)\right]^{-1}W\sum_{i\in a}P_{i}(t)$, where $W$ is the total number of walkers in the system. By replacing the sums over nodes with integrals over the activities $1/N\sum_{j}\rightarrow\int da'F(a')$ and considering the continuous limit $a$, Eq.(\ref{eq:7}) can be rewritten as:
\begin{widetext}
  \begin{align}
  \label{eq:8}
\nonumber&\frac{{\partial W_{a}(t)}}{\partial t}\\&\nonumber=-W_{a}(t)N\left\{a+mg(a)\int \frac{a^{'}F(a^{'})}{g(a^{'})+\langle g \rangle }da^{'}+m\int \frac{a^{'}g(a^{'})F(a^{'})}{g(a^{'})+\langle g \rangle }da^{'}\right\}+g(a)N\int \frac{a^{'}F(a^{'})W_{a^{'}}(t)}{g(a^{'})+\langle g \rangle }da^{'}\\&\nonumber+N\int \frac{a^{'}g(a^{'})F(a^{'})W_{a^{'}}(t)}{g(a^{'})+ \langle g \rangle }da^{'}+\frac{Namg(a)}{g(a)+ \langle g \rangle }\int F(a^{'})W_{a^{'}}(t)da^{'}+\frac{Nam}{g(a)+ \langle g \rangle }\int F(a^{'})W_{a^{'}}(t)g(a^{'})da^{'}\\&=-W_{a}(t)N\{a+\left[mg(a)\phi_{1}
+m\phi_{2}\right]\}+Ng(a)\phi_{3}+N\phi_{4}+\frac{Namg(a)\omega}{g(a)+ \langle g \rangle }+\frac{Nam}{g(a)+ \langle g \rangle }\phi_{5},
  \end{align}
\end{widetext}
where $\omega\equiv \frac{W}{N}$ is the average density of walkers per node, $\phi_{1}=\int \frac{a^{'}F(a^{'})}{g(a^{'})+\langle g \rangle }da^{'}$ and $\phi_{2}=\int \frac{a^{'}g(a^{'})F(a^{'})}{g(a^{'})+\langle g \rangle }da^{'}$ are the coefficient of $W_a$, $\phi_{3}=\int \frac{a^{'}F(a^{'})W_{a^{'}}(t)}{g(a^{'})+\langle g \rangle }da^{'}$, and $\phi_{4}=\int \frac{a^{'}g(a^{'})F(a^{'})W_{a^{'}}(t)}{g(a^{'})+ \langle g \rangle }da^{'}$ is the number of walkers that move to nodes of class $a$ due to the activation of other nodes, and $\phi_{5}=\int F(a^{'})W_{a^{'}}(t)g(a^{'})da^{'}$ is the number of walkers that move to nodes of class $a$ as a consequence of the activation. The stationary state of the process is defined by the infinite time limit $\lim_{t\rightarrow \infty}\partial W_{a}(t)/\partial t=0$. Using this condition in Eq.(\ref{eq:8}), we find the stationary solution
\begin{equation}
\label{eq:9}
W_{a}=\frac{am\omega\frac{ g(a)}{g(a)+\langle g \rangle}+g(a)\phi_{3}+\phi_{4}+\frac{am}{g(a)+\langle g \rangle }\phi_{5}}{a+mg(a)\phi_{1}+m\phi_{2}}.
\end{equation}

Hence, we can see that the quantity $W_a$ not only depends on the details of a node's activity but also on individual's memory. It is important to notice that at the stationary state $\phi_{1}$, $\phi_{2}$, $\phi_{3}$, $\phi_{4}$, and $\phi_{5}$ are constants. The values of $\phi_{3}$, $\phi_{4}$, and $\phi_{5}$ can be computed self-consistently by solving the following system of integral equations,
\begin{align}
\label{eq:10}
\nonumber& W=N\int F(a)\frac{\frac{ am\omega g(a)}{g(a)+\langle g \rangle}+g(a)\phi_{3}+\phi_{4}+\frac{am}{g(a)+\langle g \rangle }\phi_{5}}{a+mg(a)\phi_{1}+m\phi_{2}}da,\\
\nonumber&\phi_{4}=\int\frac{ag(a)F(a)}{g(a)+\langle g \rangle}\frac{\frac{am\omega g(a)}{g(a)+\langle g \rangle}+g(a)\phi_{3}+\phi_{4}+\frac{am}{g(a)+\langle g \rangle }\phi_{5}}{a+mg(a)\phi_{1}+m\phi_{2}}da,\\
&\phi_{5}=\int g(a)F(a)\frac{\frac{ am\omega g(a)}{g(a)+\langle g \rangle}+g(a)\phi_{3}+\phi_{4}+\frac{am}{g(a)+\langle g \rangle }\phi_{5}}{a+mg(a)\phi_{1}+m\phi_{2}}da.
\end{align}

For comparison, we show $W_a$ at the stationary state in the memoryless activity-driven networks as follows~\cite{perra2012random}:
\begin{equation}
\label{eq:11}
W_{a}=\frac{am\omega+\phi}{a+m\left \langle a\right \rangle},
\end{equation}
where $\omega\equiv \frac{W}{N}$ is the average density of walkers per node, $\phi=\int aF(a)W_{a}da$. We see that $W_a$ only depends on the nodes' activity.

\subsection{\label{sec:MFPT}MFPT}
We now focus on another important property of random walk process, i.e., the mean first-passage time~(MFPT), defined as the average time steps needed for a walker to visit node $i$ starting from an arbitrary node in the system~\cite{janson2012hitting}.

Let us consider $p(i,n)$ as the probability that the walker reaches the target node $i$ at time $t=n\Delta t$ for the first time. Then, $p(i,n)$ is simply given by
\begin{equation}
\label{eq:12}
 p(i,n)=\xi_{i}(1-\xi_{i})^{n-1},
 \end{equation}
where $\xi_{i}$ is the probability that the walker jumps to node $i$ during time interval $\Delta t$. The probability that a walker at node $j$ jumps to node $i$ during time $\Delta t$ is given by $\Pi_{j\to i}^{\Delta t}$~(Eq.(\ref{eq:4})). Thus, we can write $\xi_{i}$ as follows:
 \begin{equation}
 \label{eq:13}
 \xi_{i}=\sum_{j\neq i}\frac{W_{j}}{W}\Pi_{j\to i}^{\Delta t},
 \end{equation}
 where we replaced the probability that a single walker at node $j$ at time $t$ by its steady state value with $W_{j}/W$. The MFPT of node $i$ can thus be estimated as follows:
 \begin{align}
 \label{eq:14}
 \nonumber  MFPT_{i}&=\sum_{n=0}^{\infty}\Delta tnp(i,n)=\frac{\Delta t}{\xi_{i}}=\frac{\Delta t}{\sum_{j\neq i}\frac{W_{j}}{W}\prod_{j\to i}^{\Delta t}}\\\nonumber&=\frac{W}{\frac{1}{N}\sum_{j}W_{j}\left[a_{j}\frac{g(a_{i})+g(a_{j})}{g(a_{j})+\langle g \rangle}+a_{i}m\frac{g(a_{i}+g(a_{j}))}{g(a_{i}+<g>)}\right]}\\
 &=\frac{W}{g(a_{i})\phi_{3}+\phi_{4}+\frac{a_{i}mg(a_{i})}{g(a_{i})+<g>}\omega+\frac{ma_{i}}{g(a_{i})+<g>}\phi_{5}},
 \end{align}
where $\phi_3$, $\phi_4$, and $\phi_5$ are the three constants that can be calculated by Eq.(\ref{eq:10}). In numerical simulations, nodes are grouped in the same activity class $a$, which means that Eq.(\ref{eq:14}) can be written as $MFPT_a=\sum_{i\in a}MFPT_i$.

The $MFPT_i$ in memoryless activity-driven~(AD) networks is given by~\cite{perra2012random}:
\begin{align}
\label{eq:15}
 MFPT_{i}=\frac{NW}{ma_{i}W+\sum_{j}a_{j}W_{j}}.
\end{align}
We see that the MFPT obtained in the AD network model is merely determined by node's activity.

\begin{figure}[htbp]
  \centering
  \includegraphics[width=8.5cm]{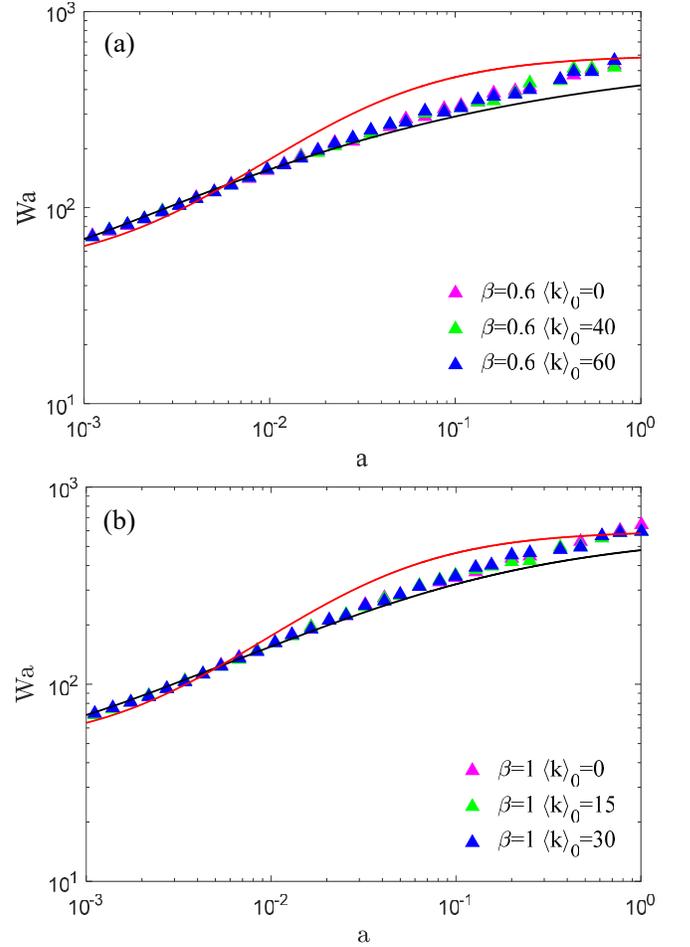}
\caption{\textbf{The distribution of walkers $W_a$ at nodes with activity $a$ in networks with different memory strengths $\beta$ and starting time, measured by $\langle k_0 \rangle$.} The red lines and black lines represent theoretical predictions without (Eq.(\ref{eq:11})) and with memory (Eq.(\ref{eq:10})), respectively. Simulation results are shown as triangles. (a) Strong memory with $\beta=1)$; (b) Weak memory with $\beta=0.6$. Other parameters are set as: $N=10^4$, $m=6$, $W/N=100$, $\gamma=2.1$. Averages are performed over $500$ independent simulations.}
\label{figure:2}
\end{figure}

\begin{figure}[htbp]
  \centering
  \includegraphics[width=8.5cm]{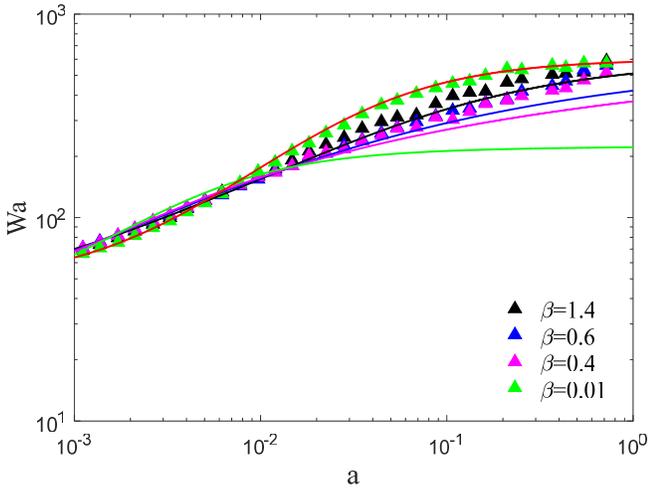}
  \caption{\textbf{The distribution of walkers $W_a$ at nodes with activity $a$ in networks for different $\beta$ with the same starting time, measured by $\langle k\rangle_0=20$.} The average number of walkers per node of class $a$ are shown as computed analytically~(continuous lines) and numerical simulations~(triangles). The red line represents the theoretical prediction of $W_a$ with Eq.(\ref{eq:11}), the other lines are predicted with Eq.(\ref{eq:10}). The parameters are set as $N=10^4$, $m=6$, $W/N=100$, $\beta=[0.01,0.4,0.6,1.4]$,  and $\gamma=2.1$. All the results are the average over $500$ independent simulations.}
\label{figure:3}
\end{figure}

\section{Results}
To support the results of the theoretical analysis, we have performed extensive Monte Carlo simulations of the random walk process on activity-driven networks with memory. We consider a power-law distribution of activity, i.e., $F(a)\sim a^{-\gamma}$, with $a\in[10^{-3},1]$. In each simulation, the temporal network evolves to time $t_0$ and then we start the random-walk process on it. We evaluate the average degree of the network at time $t_0$, as $\langle k\rangle_0$, which measures the evolution of the network at the starting time $t_0$. Networks are with size $N=10^{4}$, $m=6$, and the density of walkers is set as $\omega=10^{2}$.

Firstly, we investigate how $\langle k\rangle_0$ affects $W_a$, as shown in Fig.\ref{figure:2}. We test the effect of strong memory in Fig.\ref{figure:2}~(a) and weak memory in Fig.\ref{figure:2}~(b), respectively. For $\beta>0$, since the tail of degree distribution decays at large $k$ as $\rho(k)\propto k^{[(1+\beta)\gamma+1]}$~\cite{ubaldi2016asymptotic}, it results in a decrease of $W_a$, compared to that for the memoryless case. More in details, depending on the activity value $``a"$, memory shows different effects on the distribution of $W_a$, thus showing a crossing point among the curves. For smaller activity $a$, memory helps nodes collect more walkers than that in memoryless case. This is due to that, affected by individual's memory, it is difficult for the nodes with smaller activity to receive links in the network evolution. Therefore, nodes with smaller activity release less walkers but receive more. With the increase of activity $a$, the stronger the memory is, the lower the $W_a$ will be. This is because frequent sending and receiving links reduce the number of collected walkers, resulting in that the degree distribution decays at large $k$.

In the following, we mainly demonstrate the effect of memory for the nodes with large activity. Regardless of the weak memory ($\beta=0.6$) or the strong memory ($\beta=1$), although the connectivity of starting network, $\langle k\rangle_0$, is different, the steady state of random-walk process is rarely affected. Since the evolution time for the steady-state of random-walk process is larger than the start time of the network evolution, individual's memory dominates the network evolution.

Then, we study how memory strength $\beta$ affects $W_a$ by keeping $\langle k\rangle_0=20, \gamma=2.1$ fixed in Fig.\ref{figure:3}. We see that for $\beta>0.01$, the larger $\beta$ is, the stronger the nodes' ability to collect walkers will be, which has been verified in the theoretical analysis (see Fig.\ref{figure:3} solid lines). Surprisingly, with the increase of $\beta$, $W_a$ approaches the value of $W_a$ in the memoryless case. For the same reason as explained above, for $\beta>0$, $\rho(k)\propto k^{[(1+\beta)\gamma+1]}$, the difference of $W_a$ between the networks with memory and without memory reaches the maximum in the limit $\beta\rightarrow 0$, since degree inhomogeneities are stronger in this case. For small values of $\langle k\rangle_0$, since the creation of new links is not neglectable, simulation results are higher than theoretical results. For $\beta=0.01$, the creation of new links dominates the dynamics and it increases the number of walkers at nodes with large activity towards the memoryless case.

\begin{figure}
  \centering
\centering
\includegraphics[width=8.5cm]{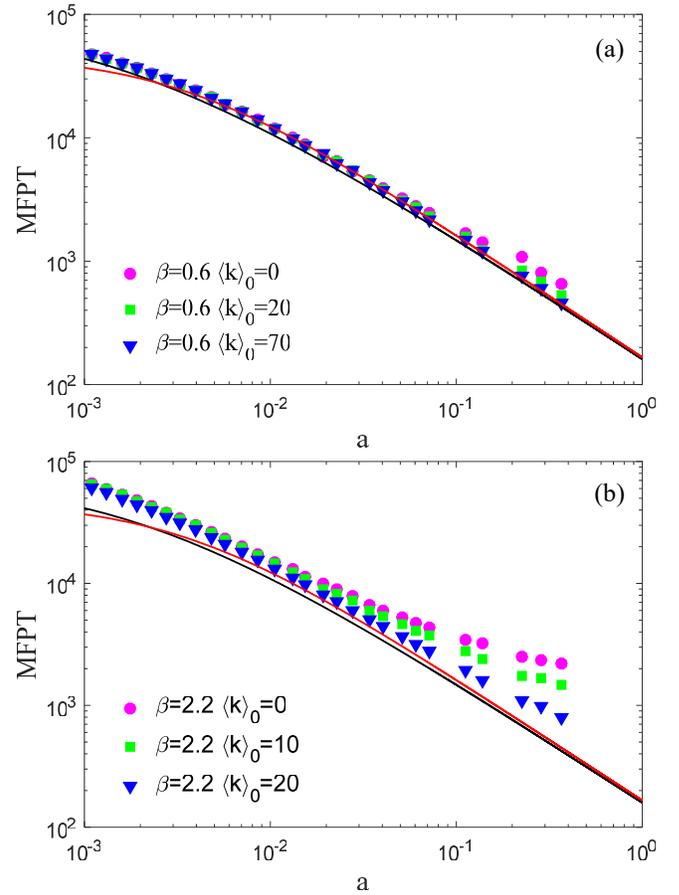}
\caption{\textbf{The MFPT versus activity $a$ for different $\langle k\rangle_0$.} (a) $\langle k\rangle_0=[0,20,70]$; (b) $\langle k\rangle_0=[0,10,20]$. The red lines and black lines represent theoretical predictions without (Eq.(\ref{eq:15})) and with memory (Eq.(\ref{eq:14})), respectively. Simulation results are shown as symbols~(dots, squares, and triangles). The network size is $N=10^3$, $m=6$, $\gamma=2$. Each point is the average over $10^3$ independent simulations.}
\label{figure:4}
\end{figure}

\begin{figure}[ht]
  \centering
  \includegraphics[width=8.5cm]{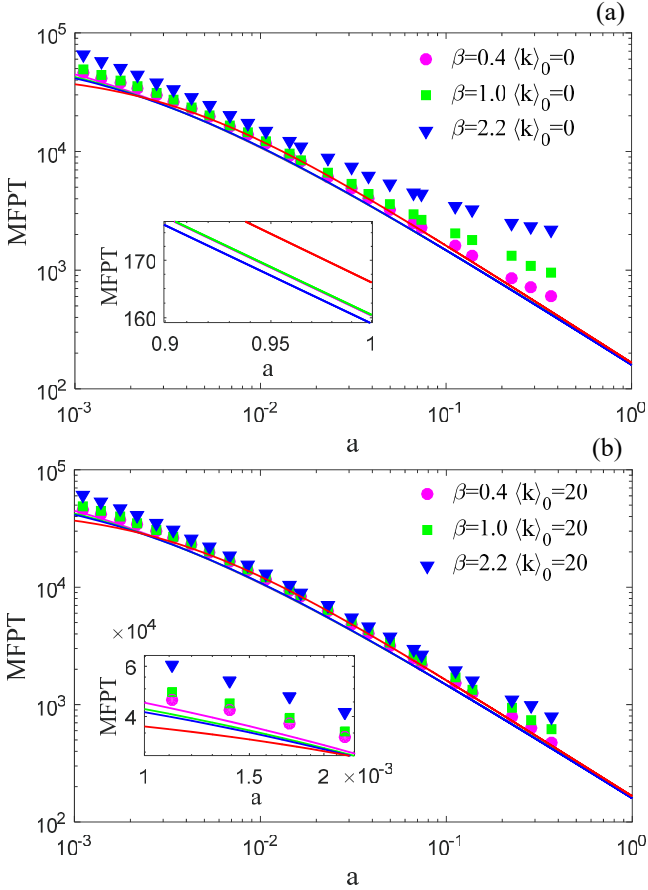}
  \caption{\textbf{The MFPT versus activity $a$ for different $\beta$.} The inner figure is the enlarged figure of the cure on the left and right, respectively. (a) $\langle k\rangle_0=0$; (b) $\langle k\rangle_0=20$. Solid lines represent analytically results (Eq.(\ref{eq:14}) and Eq.(\ref{eq:15})) and symbols represent simulations for different values of $\beta$. The parameters are set as $N=10^3$, $m=6$, $\gamma=2$, $\langle k\rangle_0=20$, $W=1$, $\beta=[0.4,1.0,2.2]$. Each point is the average result over $10^3$ times.}
 \label{figure:5}
\end{figure}

\subsection{MFPT}
Next, we verify the theoretical results of MFPT by setting the parameters as follows: $N=10^3$, $m=6$, $W=1$, and $\gamma=2.1$. The results of MFPT for numerical simulations are compared with theoretical predictions for different choices of memory strength $\beta$ and starting time, measured by $\langle k \rangle_0$ in Fig.\ref{figure:4}. With a weak memory strength $\beta=0.6$~(Fig.\ref{figure:4}~(a)), we find that simulation results converge toward the analytical prediction as $\langle k\rangle_0$ increases, which is more obvious when memory strength is strong as shown in ~(Fig.\ref{figure:4}~(b)). When $\langle k\rangle_0$ is small, the creation of new links dominates network evolution, leading to a strong deviation from the theoretical prediction~(Fig.\ref{figure:4} black line). The reinforcement mechanism of edges leads to the formation of cluster-like structure in the network, which means that the walker can easily get trapped at the traversed nodes and hardly jump to new nodes. Individual's memory can be observed to slow down the transport dynamics, as we observe that in this case the MFPT is larger for all the nodes compared to the memoryless case~(red line), as shown in Fig.\ref{figure:4}. For larger $\beta$, the strengthening mechanisms between edges become even more significant, which further delays the MFPT.

Then, we further study the interplay of memory strength $\beta$ and $\langle k \rangle_0$ on the MFPT of random-walk process in details in Fig.\ref{figure:5}. To clarify the impact of $\langle k \rangle_0$, we testify typical cases. With an extremely small value $\langle k \rangle_0=0$, it represents a special case that the starting time of network evolution and random walk are same. \textbf{} With a larger value of $\langle k \rangle_0=20$, \textbf{}it enhances the memory effect in the network. However, due to the constraint of simulation, the larger $\beta$ is, the more difficult it is to generate networks with larger $\langle k \rangle_0$. We achieve the network evolution with a relatively small value of $\langle k \rangle_0=20$ under different $\beta$. Similar to $W_a$, the theoretical curve of MFPT also shows a crossover phenomenon, that is, we also have to discuss separately from the case of small activity and large activity. In theory(solid line in the subgraph of Fig.\ref{figure:5}(b)), when the activity $<2\times10^{-3}$, the stronger the memory, the fewer links the node receive, and it takes a long time for walkers to reach the nodes. The nodes with large activity indeed has large degree and can form clusters where the high frequency of mutual contacts allow for reinfections and positive correlations, which allows walkers to arrive quickly (shown as Fig.\ref{figure:5}(a)).
However, the simulation results show the opposite results, as for small $\langle k \rangle_0$, there is a gap between simulation results and the theory. Obviously, we can see that as $\langle k \rangle_0$ increases, the gap between simulation and analytical results decrease. However, we expect that for large enough $\langle k \rangle_0$, at any $\beta>0$, the dynamics is dominated by individuals' memory and the MFPT recovers the analytical results.

It can be seen from Fig.\ref{figure:4} and Fig.\ref{figure:5}, the larger the activity of a node is, the less the average time it takes for the walker to reach it. It shows that during the process of network evolution, nodes with large activity are frequently activated and attract walkers to jump on them quickly.

\section{\label{RealWR}Random walks process in real networks}
We further investigate how individual's memory affects the random-walk process in real networks. We collect the interactions containing time-stamped information between $30398$ Digg users in August 2008 via 87627 reply network~\cite{de2009social}. The Digg-Reply data is time-varying, where each node describes a user and each time-resolved link denotes that a user replied to another user. Since many users tend to interact with the users in same group for multiple times, the social network is obviously driven by non-Markovian human dynamics.

\begin{figure}[htbp]
  \centering
  \includegraphics[width=8cm]{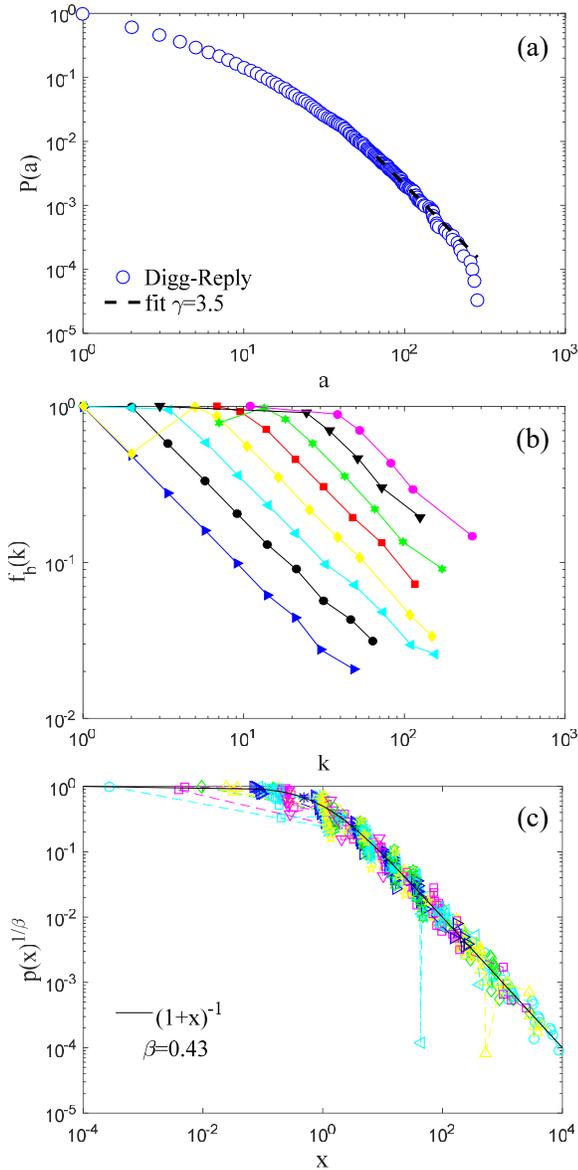}
  \caption{\textbf{Statistical properties in Digg-Reply dataset.} (a) Cumulative activity distribution $P(a)$ in the Digg-Reply dataset. The dashed line represent the fitting result of  power-law distribution with exponent $\gamma=3.5$; (b) The probability that nodes establish a new connection as a function of their degree. Each data sequence~(different colours and markers) corresponds to a selected nodes of the system, with the average activity of the class increasing from the bottom curves to the upper curves; (c) We rescale the attachment rate curves of all the nodes by setting $k\rightarrow x_b=k/c_b$ and plot $p_b(x_b)^{1/\beta}$ versus $x$, where $\beta$ is same for each same colored curve. The memory strength is fitted as $\beta =0.43$. }
\label{figure:6}
\end{figure}

In order to characterise individual's memory in the network, we measure the activity $a_{i}$, defined as the fraction of interactions of node $i$ per unit of time, which describes the propensity of node $i$ to be involved in social interactions, is computed as $a_i=s_{i,out}\sum_js_{j,out}$, where $s_{i,out}$ is the out-strength of node $i$ integrated across the entire time span~\cite{Alessandretti2017Random}. Then, we use a binning method to divide the nodes in total number, $N_{b}=\sum_{a=1}^{N_{act}}N_{deg}(a)$, of activity-degree classes according to their activity $a$ and final degree $k$~\cite{ubaldi2016asymptotic}, i.e., nodes that engaged a similar number of interactions and that have a comparable cumulative degree in the observation period. We define $e_{b}(k)$ as the total number of events engaged by the nodes of the $b-$th class with degree $k$, and $n_{b}(k)$ as the total number of events that the nodes belonging to the $b-$th and featuring degree $k$ perform toward a new node.  We measure the reinforcement process in the Digg reply network by minimizing the function $\chi^2{(\beta)}$~\cite{ubaldi2016asymptotic}:
\begin{figure}[htbp]
  \centering
  \includegraphics[width=9cm]{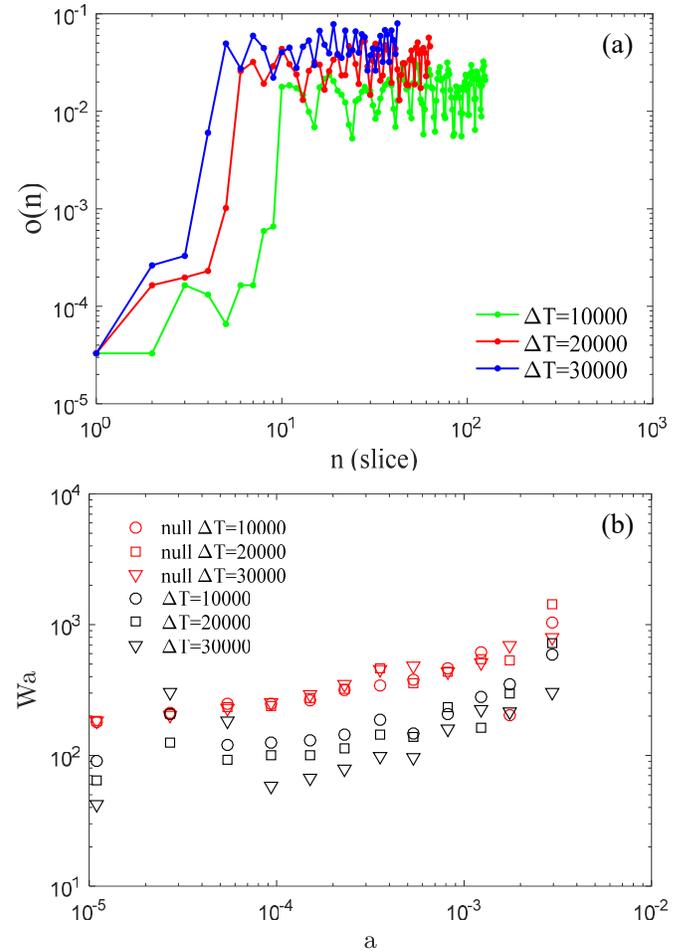}
\caption{\textbf{Random-walk process in Digg-Reply dataset.} (a) The fraction of active nodes for different choices of time-slice $\Delta T$; (b) The distribution of $W_a$ in the Digg dataset~(black) and in the null model (red) for different window size $\Delta T=[10000,20000,30000]$ with $W/N=200$. Each point is the average over $2000$ independent simulations.}
 \label{figure:7}
\end{figure}
 \begin{align}
 \label{eq16}
 \chi^2{(\beta)}&=\sum_{b=1}^{N_{b}}\chi_{b}^2{(\beta)}=\sum_{b=1}^{N_{b}}\sum_{k=1}^{k_{b}}
 \frac{\left[f_{b}(k)-p_{b}(k,\beta)\right]^2}{\sigma_{b}(k)^2},
 \end{align}
where $f_{b}(k)=n_{b}(k)/e_{b}(k)$ is the probability for a node with degree $k$ to get a new connection. $p_{b}(k,\beta)=(1+\frac{k}{c(b)})^{-\beta}$ is a reinforcement function used to fit $f_{b}(k)$. $\sigma_{b}(k)=\sqrt{\frac{f_b(k)(1-f_b(k)}{e_b(k)}}$ is the STD of $f_b(k)$.

In Fig.\ref{figure:6}~(a), we performed a power-law distribution fitted on the Digg-Reply data set, which shows that the fluctuation of nodes' activity is small fitted with the exponent  $\gamma=3.5$. Consider that nodes with the same activity may feature different memory behavior, for example, individuals with large activity may connect to very few different nodes (strong memory) or establish new links at almost every step (weak memory). For this reason, we group the original data of Digg-Reply according to nodes' activity and degrees, and calculate the probability that a new edge connects with a new node with degree $k$ in each category of nodes' activity, as shown Fig.\ref{figure:6}~(b). For curves with small activity category (bottom curves), the probability to attach to a new node quickly drops to $0$ with degree $k \lesssim10$. However, for large activity category (top curves), even with very large degree $(k\sim10^2)$, the probability is nontrivial with $p(k)\gtrsim0.1$. As can be seen in Fig.~\ref{figure:6}(c), an obvious memory effect in the Digg reply dataset is observed. It is well fitted with a single memory strength, e.g., $\beta_{opt}=0.43$. When the number of individual's acquaintances is larger, the probability of connecting with a new node is small, thus we see that $p_b(k)$ decreases with degree $k$.

In order to compare the effect of memory, we need to randomize the network to remove the memory effect and take it as the null network. Randomization is performed by recombining the interactions at each timestamp in order to remove the memory effect, while retaining the order of activation time for each node, the final time integration degree distribution, and the degree distribution at each time step~\cite{starnini2012random}.

To explore the effect of the number of active nodes on random walks, we divide the original data according to the time interval $\Delta T=10000,\Delta T=20000$, and $\Delta T=30000$, respectively. As can be seen from Fig.\ref{figure:7}~(a), with the increase of time interval $\Delta T$, the number of active nodes in the time accumulated network increases accordingly. In Fig.\ref{figure:7}~(b), we compare the distributions of walkers $W_a$ versus activity $a$ on the data network with the null network with different time interval $\Delta T$. In the null model, since the randomization process eleminates the effect of memory, $W_a$ is hardly affected by $\Delta T$. Compared with the null model, strong ties established by memory in the real data lower $W_a$ for all the time slices $\Delta T$ we tested, which are consistent with our results on artical networks. Due to the lower heterogeneity of node's activity in Digg-Reply network~(Fig.\ref{figure:6}~(a) with $\gamma =3.5$), the number of walkers fluctuates less with node's activity, thus, we see a flat increase in $W_a$.

\section{\label{Discussion}Discussion and conclusion}
In this work, we investigate the random walk process on temporal networks with memory. We study how individual's memory and the starting time of the diffusion co-affect random walk process unfolding on the network. Under the constraints of long-time evolution, we derived analytical expressions of the stationary state and the distribution of mean first-passage time.

Monte Carlo simulation results show that, compared with the memoryless case, individual's memory enhances degree heterogeneity. For smaller $\beta$, even if the random-walk process starts at large $\langle k\rangle_0$, the creation of new links dominates the dynamics, which is equivalent to the memoryless case. For larger $\beta$, the strengthening mechanisms between edges becomes even more significant, making the walker trapped at traversed nodes and delay the MFPT. $\langle k\rangle_0$ plays a trivial effect on $W_a$ and greatly affects the MFPT. Numerical results show that the MFPT converges toward the analytical prediction as $\langle k\rangle_0$ becomes large.

We perform similar analysis on a real networks, i.e., Digg-Reply dataset. The network shows obvious memory effect with $\beta=0.43$. Compared with the null model of Digg-Reply, we find that individual's memory limits the nodes' capability of gathering walkers, which is consistent with the results of the artificial networks.

In conclusion, our work provides a comprehensive view for the random-walk process in temporal networks with memory, compared to that in memoryless case. In the presence of memory, the number of walkers decreases at steady state and the MFPT gets larger than that of the memoryless case. Moreover, the effect of individual's memory on the random-walks in real data verifies the results on the artificial networks. As a possible future work, memory can be incoroperated in time-varying networks with individual's higher-order interaction.

\section*{Acknowledgements}
This work was supported by the National Key Research and Development Program of China under Grant No.~2017YFE0117500, the National Natural Science Foundation of China under Grant No. 61603237, and the Program for Professor of Special Appointment (Eastern Scholar) at Shanghai Institutions of Higher Learning.

\bibliography{ref}% Produces the bibliography via BibTeX.
\bibliographystyle{plain}
\end{document}